# THERMAL HISTORY OF $CB_b$ CHONDRULES AND COOLING RATE DISTRIBUTIONS OF EJECTA PLUMES.

Running head: Cooling of Chondrules and Ejecta Plumes

R. H. Hewins[1,2], C. Condie[1,3], M. Morris[4], M.L.A. Richardson[5], N. Ouellette[4], M. Metcalf[4].

[1]EPS, Rutgers University, Piscataway NJ 08816, USA
[2]IMPMC, MNHN, UPMC, Sorbonne Universités, Paris 75005, France
[3]Natural Science, Middlesex Community College, Edison, NJ 08818, USA
[4]Physics, SUNY at Cortland, NY 13045, USA
[5]Sub-department of Astrophysics, University of Oxford, Keble Road, OX1 3RH, UK
*2018 January 16th*

## ABSTRACT

It has been proposed that some meteorites, CB and CH chondrites, contain material formed as a result of a protoplanetary collision during accretion. Their melt droplets (chondrules) and FeNi metal are proposed to have formed by evaporation and condensation in the resulting impact plume. We observe that the SO (skeletal olivine) chondrules in $CB_b$ chondrites have a blebby texture and an enrichment in refractory elements not found in normal chondrules. Since the texture requires complete melting, their maximum liquidus temperature 1928 K represents a minimum temperature for the putative plume. Dynamic crystallization experiments show that the SO texture can be created only by brief reheating episodes during crystallization giving partial dissolution of olivine. The ejecta plume formed in a smoothed particle hydrodynamics (SPH) simulation (Asphaug et al., 2011) served as the basis for 3D modeling with the adaptive mesh refinement (AMR) code `FLASH4.3`. Tracer particles that move with the fluid cells are used to measure the *in situ* cooling rates. Their cooling rates are ~10,000K/hr briefly at peak temperature and, in the densest regions of the plume, ~100 K/hr for 1400-1600 K. A small fraction of cells is seen to be heating at any one time, with heating spikes explained by compression of parcels of gas in a heterogeneous patchy plume. These temperature fluctuations are comparable to those required in crystallization experiments. For the first time, we find agreement between experiment and models that supports the plume model specifically for the formation of CBb chondrules.

1. INTRODUCTION

The CR carbonaceous chondrite clan consists of CR, $CB_a$, $CB_b$ and CH chondrites (Weisberg et al., 1995; Krot et al., 2002). Unlike the CR chondrites, the other groups are metal-rich meteorites and contain chondrules differing in texture and composition from 'normal' chondrules (Weisberg et al., 1988; Krot et al., 2001a, 2005). Their chondrules and FeNi metal show volatility-related fractionations and have an age of $4.8 \pm 0.3$ Myr after refractory inclusions younger than 'normal' chondrules ~0-3 Myr after CAI (Krot et al., 2005; Meibom et al., 1999; Kleine et al., 2005; Bollard et al., 2015). This indicates a late event in a gas-poor disk. A consensus is developing for an origin of the 'non-normal' chondrules by melting, evaporation and condensation in an impact plume (Krot et al., 2005) and the physics of collisions, jetting, and ejecta transport is being studied (Asphaug, 2011; Dullemond, 2014; Morris et al., 2015; Johnson et al., 2015; Hasegawa et al., 2016). Here we have developed two tests of this concept: the determination of the thermal history of the chondrules (1) using dynamic crystallization experiments on $CB_b$ chondrule compositions, (2) using models of impact plumes with the Eulerian AMR (Adaptive Mesh Refinement) code, FLASH4.3 (Fryxell et al. 2000). Note that Lagrangian Smoothed Particle Hydrodynamics (SPH) discretizes the fluid into individual parcels of material that act as an interpolant for determining the physical characteristics at a set location. Previous SPH work of impacts and their plumes have at times individual particles being ejected, with the pressure essential set to zero. In some regions this means the ejecta plume is not well resolved, and variations in pressure cannot be studied. Here we attempt to overcome these issues by converting an SPH dataset early in its plume's evolution to an AMR scheme. We find concordance of the thermal histories obtained by experiment and models, providing a specific demonstration of the viability of the chondrule impact plume origin.

## 2. CONSTRAINTS FROM NATURAL CHONDRULES

In Figure 1, we show the compositions of SO and CC (cryptocrystalline) chondrules (Krot et al., 2002, 2010), supplemented by our own Hammada al Hamra 237 data measured using a Cameca SX100 EMP and similar procedures to Hewins et al. (2012). These chondrules are depleted in moderately volatile elements, and show volatility-controlled composition trends (Figure 1a,b). They have been interpreted as evaporation residues and condensates, respectively, in a vapor-melt impact plume (Krot et al., 2005).

It is well known (Lofgren, 1996; and references therein) that complete destruction of nuclei above critical size during melting is necessary for the growth of skeletal and dendritic crystals on cooling. Crystal morphology is independent of heating rate, being controlled by the nature of potential nucleation sites or embryos (Lofgren, 1966, Hewins et al., 2005). Because the texture of the SO evaporation residues requires complete melting, the highest calculated liquidus temperature of the SO chondrules gives a minimum estimate of the impact plume temperature, assuming that the chondrules move along with the fluid frame with no relative motion and no extra frictional heating. Figure 1c shows that this is 1655°C (1928K). We do not consider CC chondrules here, as such liquids can condense either above or metastably below their liquidus temperatures, giving no constraint on ambient conditions. $CB_b$ zoned metal enclosing CC chondrules condensed only after cooling to 1375K (Campbell et al., 2005). The 1928K minimum is compatible with the amount of evaporation expected at model peak temperatures of 2000K (Dullemond et al., 2014; Johnson et al., 2015) and 2200K (this work).

Because of CR chondrite affinities with CH and CB material, we considered the CR Renazzo silicate fraction (in a chondritic or differentiated body) as a possible source of the SO chondrules, but it is too FeO- and Na-rich (Jarosewich, 1990). Renazzo silicate and SO chondrule compositions would be superheated if totally liquid at plume peak temperature (Figure 1c). Boiling and evaporation of the moderately refractory elements is limited by the very high initial cooling rates, but evaporation would accompany eventual crystallization. For liquid like Renazzo silicate there would be rapid loss of FeO, Na, etc. by boiling during expansion (cf. Krot et al., 2005). We calculated a Renazzo silicate composition with FeO reduced to 3%, as in SO chondrules. Figure 1 shows that this composition plots near the most Al-poor SO chondrules and could be the immediate parent from which less volatile elements were lost. $Al_2O_3$ and CaO are seen to be enriched four-fold over the putative parent composition, with a corresponding drop in MgO normalized to initial MgO (Figure 1a,b). The enrichment of Al and Ca can be achieved by evaporative loss of ~30% of the MgO and ~15% of the $SiO_2$ in an FeO-poor parent.

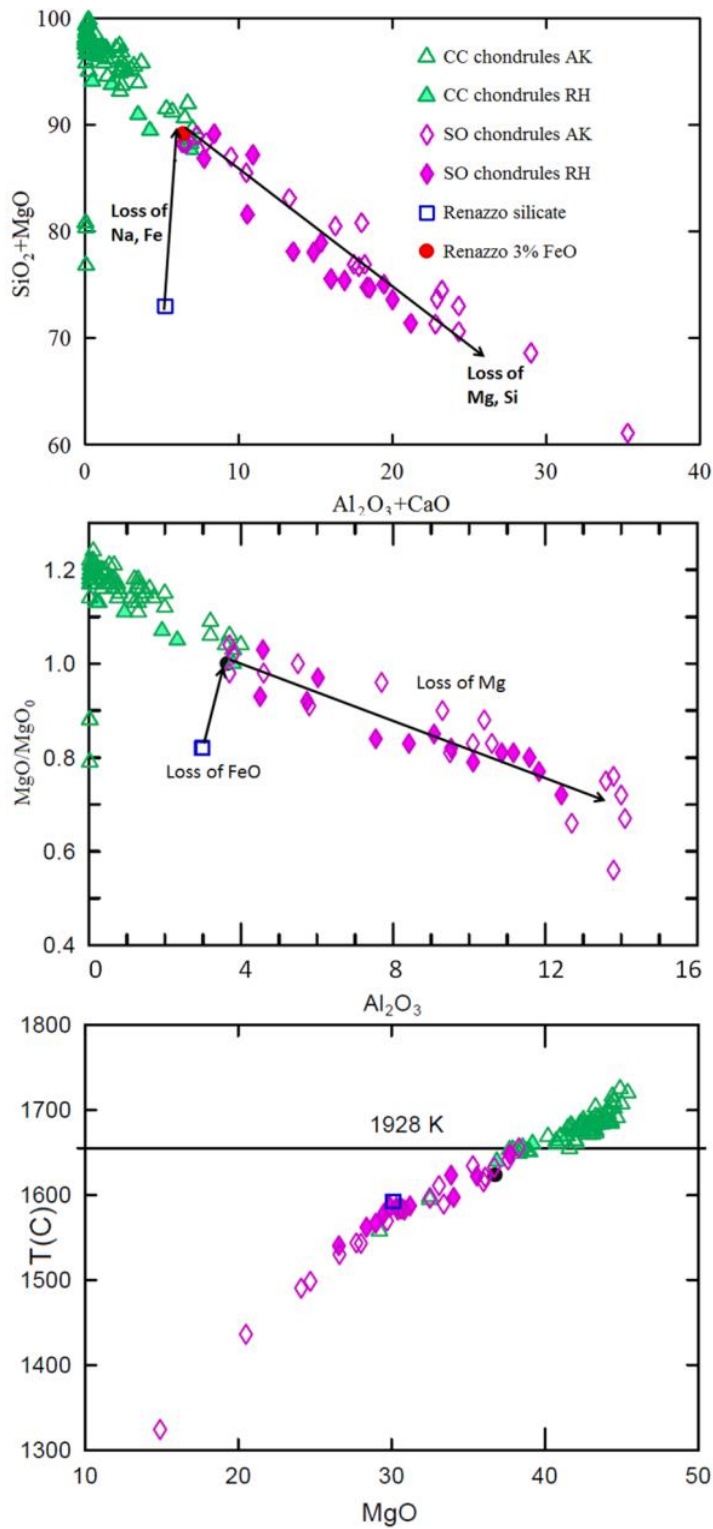

Figure 1 Chondrules in $CB_b$ chondrites and Isheyevo (Krot et al., 2002, 2010; this work) (a,b) Correlation of highly versus moderately refractory elements indicate volatility-controlled compositions. SO chondrules can be explained by evaporative loss of Mg and Si from an FeO-depleted CR silicate composition. (c) The maximum calculated liquidus temperature for SO chondrules gives a minimum value for the ambient temperature.

## 3. COOLING HISTORY FROM EXPERIMENTAL PETROLOGY

Donaldson (1976) and Faure et al. (2003, 2007) have studied the morphology of olivine crystals as a function of degree of undercooling and cooling rate. Olivine in experiments with ‒normalø chondrule compositions has also been widely studied (e.g. Connolly et al., 1998; Hewins et al., 2005) and shows a hierarchy of olivine growth habits, but without the skeletal blebby olivine (SO) chondrules typical of $CB_b$ chondrites. This texture resembles barred (parallel plate) olivine, but is distinct from it with parallel sets of discontinuous rounded blebs instead of faceted bars (Figure 2a,b).

Condie (2012) did dynamic crystallization experiments in a Deltech Dt-31-VT-OS muffle tube furnace on Al-rich compositions close to those of chondrules SBO5 and SBO6 of Krot et al. (Table 4, 2002). Techniques similar to those of Fox and Hewins (2005) were used, with super-liquidus peak heating temperatures, a wide range of heating times and quench temperatures, and cooling rates of 10-1000°C/hr. The superliquidus temperatures excluded the formation of porphyritic olivine chondrules, which are not present in $CB_b$ chondrites. These runs produced a series of hopper, linked parallel growth and dendritic rod olivine crystals, similar to those described as elongate hopper and barred olivine by Connolly et al. (1998). All of these olivine habits are characteristic of rapid growth (Faure et al., 2007) and unlike those of $CB_b$ SO chondrule olivine: the SO chondrule texture cannot be due to the high Al and Ca concentrations.

As the skeletal blebby texture suggested annealing of barred olivine (cf. Tsuchiyama et al., 2004) we performed a series of runs in which cooling was arrested at ~200°C below the liquidus, where the charge soaked at 1300°C for 4 days. We observed little or no change to the sharply faceted olivine. We therefore performed similar runs in which the soak was interrupted several times at one day intervals and the charge briefly reheated over 40-72 minutes to 1420-1515°C, before final cooling at 1000°C/hr and quenching. Reheating would lead to partial dissolution of the olivine bars, probably in much less than the one day soak times used. This approach was successful for runs on SO5 (Figure 2c), with rounded blebby olivine grains and a close texture match to natural SO chondrules. The same reheating program failed for the less refractory SO6 composition because the late temperature spikes equaled or exceeded the liquidus temperature.

Although this texture has not been obtained in ‒normalø chondrule experiments, it was found in Na partitioning experiments, where olivine crystals were equilibrated with a CMAS silicate melt at 1300°C (Kropf, 2009). These charges were given heating spikes of 15°C at 12 hour intervals. Thus the blebby skeletal olivine texture also found in SO chondrules requires temperature fluctuations during cooling. Cooling rates producing the presumed parent barred olivine texture are 100-1000°C/hr (Radomsky and Hewins, 1990; Tsuchiyama et al., 2004), and therefore a wide range of cooling rates could have given rise to SO chondrules.

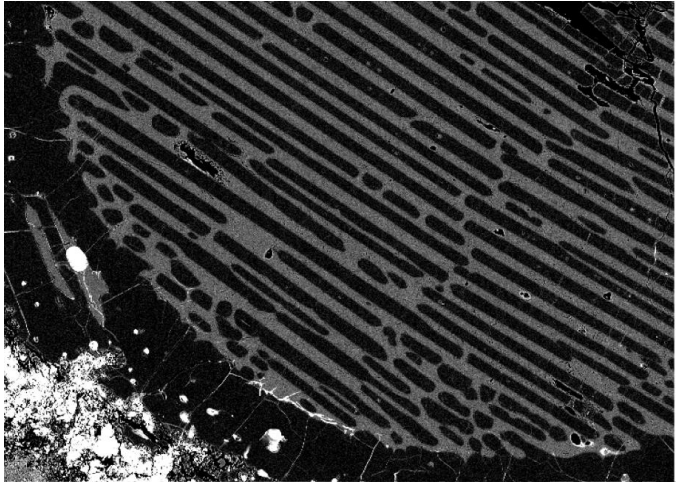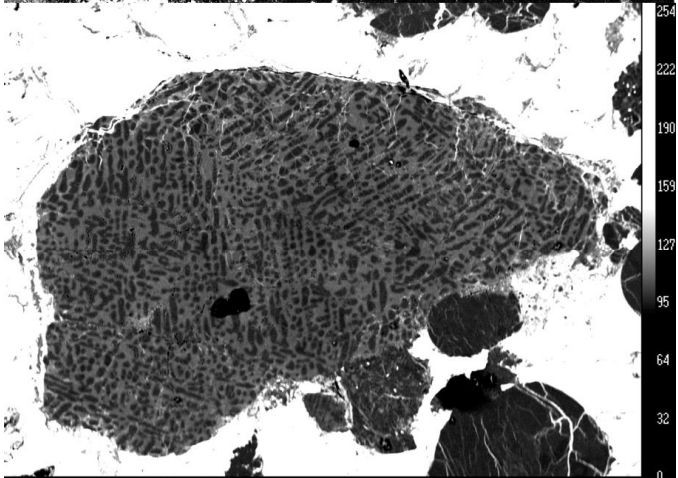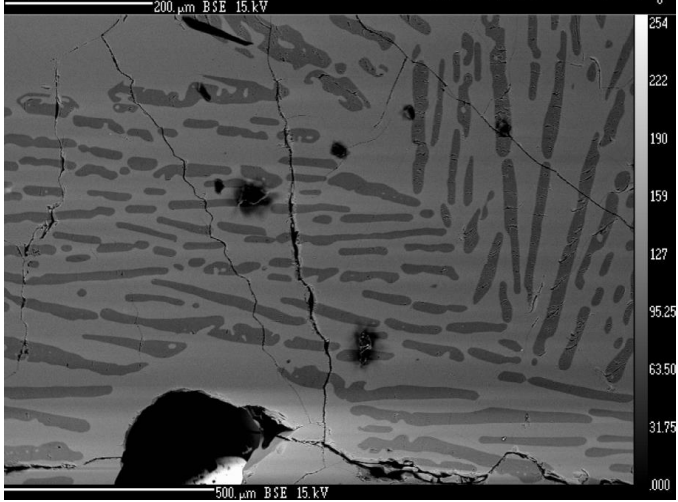

Figure 2. BSE images demonstrating skeletal olivine texture. (a) A "normal" barred olivine chondrule in the Kaba CV3 chondrite. (b) An SO chondrule in Hammadah al Hamra $CB_b$ chondrite. (c) Skeletal olivine texture in charge SBO5-20, reheated during cooling.

## 4. NUMERICAL METHODS

We begin with the 3D model from Asphaug et al. (2011), a smoothed particle hydrodynamics (SPH) simulation of an ejecta plume after a collision of 30 and 100 km radius basaltic planetesimals at twice the escape speed, $v_{esc} \sim 110$ m/s. The ejected material has its pressure set to zero and evolves ballistically, allowing the model to adiabatically expand, consistent with a pressure-bound vapor phase with sub-resolution sized melt droplets. Note that Asphaug et al. (2011) do not model the vapor phase and its pressure variation, and cannot then describe the thermal evolution of the material that forms droplets. Using the method of Richardson et al. (2013) we map the dataset 4000 s after collision into the adaptive mesh refinement (AMR) code `FLASH4.3` of Fryxell et al. (2011). For this work, we use a maximum resolution of 1 km, slightly better than the minimum smoothing length in the SPH simulation of the plume. We can then (i) model the shock heating of the nebular medium as the plume expands; (ii) implement a radiative transfer model so droplets in the plume may radiate and cool, accounting for the opacity in the plume and nebula; iii) fix the spatial resolution as the plume expands.

We include the interaction between the nebula and the plume, and also allow the plume to evolve under self-gravity. Thus, we set the initial pressure in our model (both in the mapped material and in the ambient nebula) to a constant value, consistent with a pressure-bound plume that is mostly a pure ideal-gas vapor. As the plume expands it shock-heats the nebula, and any collapse or expansion has a resulting change in pressure. We assume a uniform initial temperature of $T_o = 2200$ K, consistent with a completely vaporized plume if equilibrium were achieved. Note our results are insensitive to this temperature choice, as only the magnitude of any cooling and heating would scale with $T_o$, therefore the qualitative results are robust. The plume cools adiabatically as it expands. We modeled the cooling with a mass scalar that advects with the flow. We initialize this mass scalar to the inverse of the local density, such that at later times the value of the mass scalar in a cell is a mass-weighted average of the initial density of all gas that has merged into this cell. Neglecting the melt, and assuming the vapor is an ideal gas, we have:

$$T(\mathbf{x},t) = T_o[\rho(\mathbf{x},t)/\rho_o(\mathbf{x},t)]^{\gamma-1} = T_o[\rho(\mathbf{x},t)X(\mathbf{x},t)]^{\gamma-1},$$

where the temperature, $T$, is found at every point in space, $\mathbf{x}$, and time, $t$, as a function of the initial temperature, the ratio of the specific heats, $\gamma = 1.4$, the local density, $\rho$, and the mass scalar, $X$, which is the inverse of the initial density, $\rho_o$. We assume the plume is a vapor-melt mix, where sub-mm droplets of mean density 2.7 g/cm$^3$ and atomic weight 64.5 g/mol are in thermal equilibrium with the lower density vapor. While the validity of this assumption is likely poor, it would set a maximum cooling and heating rate for the melt when radiative cooling is not included.

To measure the evolving cooling rates in the bulk vapor requires introducing tracer particles that move with the fluid. With these particles, we can sample the temperature of regions of the plume as it moves, giving a comoving thermal history. Tracer particles are massless, and are only used as a tool to follow plume parcels to estimate cooling rates of comoving melts that are otherwise unresolved in the simulation. These particles are initialized randomly, following a probability distribution given by the mass fraction in a cell. Unfortunately, unlike SPH, which is a pure Lagrangian method, tracer particles implemented in an Eulerian code will not stay perfectly within the same parcel of gas. We account for times when the tracer particle slips out of (and possibly back into) its parcel by discounting a time step if it has a sudden change in the cooling rate (second derivative of temperature) larger than 100 K/hr/s. We calculate cooling rates from the change in temperature over the remaining time steps. We further test the validity of these calculated cooling and heating rates by measuring the parcel entropy, and the cooling rates in the neighboring 500 cells. Given our assumption of adiabatic evolution, the entropy should be roughly constant, only changing slowly as mixing of material occurs. Further, the cooling rate of the tracer should match the average cooling/heating rates of surrounding cells. Provided these two requirements are met, we assume this cooling / heating rate tracked by the tracer is accurate.

We have run a collection of 3D simulations (run on the Extreme Science and Engineering Discovery Environment (XSEDE); Towns et al. 2014), varying the resolution, ambient density, the initial pressure, the self-gravity, and radiative cooling. The results from this parameter suite are presented in M. L. A. Richardson et al. (2017). For the discussion here, we only focus on our high-resolution simulation, where the nebula density, $\rho_N$, is $10^{-11}$ g cm$^{-3}$. The initial pressure is 1.29 x 10$^{-9}$ bar, the resolution is 1 km, radiative cooling is off (acceptable for early evolution before the plume becomes optically thin), and no gravity from the remnant body is included.

## 5. MODEL RESULTS

The ejecta plume has a mass of 1.23x10$^{20}$ g and begins with a uniform temperature of 2,200 K, higher than the minimum peak temperature indicated by SO chondrules. If CB$_a$ metal represents metal coexisting with Fe-rich vapor at the beginning of expansion, the temperature could be as high as 2500 K (Campbell et al., 2002). The plume extends about 1,500 km, with a typical expansion speed of 1000 km/hr. We used nearly 10,000 tracer particles to trace the temperature evolution of the plume for 5000s. Projections through the plume are provided in Figures 3a,b,c taken at 2, 40, and 83 minutes. We also include the tracer particle positions, with colors corresponding to the temperature of the vapor they are tracing. Tracers in gas above 1800 K are red, tracers in cooler gas above 1600 K are orange, tracers in yet cooler gas above 1400 K are yellow, and tracers in even cooler gas are blue.

Using the density and temperature information of the tracer particles, we identify spatial regions of the plume with consistent temperature evolution. We find that regions near the top and bottom of the plume cool the fastest, while heating of the nebula occurs along the front of plume. Most of the plume (by mass) has cooled to ~1,400 K and expanded by 1,000 km after 1 hour (see Figure 3c). However, due to variations in the initial velocity field, there are similarly significant variations in the density and temperature evolution across the whole plume. A single fluid parcel will cool at different rates throughout its evolution, though, in general, cooling rates are seen to decrease with decreasing temperature. These results are robust across our whole parameter suite, although we only focus here on our high resolution run.

We present two snapshots of the plume's thermal evolution at roughly two minutes and 40 minutes (Figure 4a,b). Most of the fluid parcels in the plume are first superheated, and by 2 minutes are cooling at moderately fast rates (~1500 K/hr), though a few are cooling from 6,000 - 10,000 K/hr. By 40 minutes some of the plume has cooled below our temperature range, to pass the glass transition temperature (~1,100K; Wick and Jones, 2012), and most of the plume is cooling at roughly 1,000 K/hr in the temperature range 1400 - 1700 K. Regions that would form chondrules are concentrated in the high density regions and must crystallize while cooling at "low" rates (0-300 K/hr), but those cooling at the highest rates would be quenched, becoming either glassy or cryptocrystalline.

A small fraction of cells is seen to be heating at any given time (Figure 4), though their general history is, of course, one of cooling. Cooling paths for two selected tracer particles are shown in Figure 5. Their cooling rates are ~1,300K/hr or less down to a temperature of ~1,900K, and subsequently 300-400K/hr. Overall, we find a variety of average cooling rates, with the highest ones above 2500 K/hr occurring at the periphery of the plume, at intermediate densities. The densest regions of the plume have characteristically lower cooling rates, ~100 K/hr for 1400-1600 K. At lower initial densities the plume is more efficient at cooling, with 19% (by mass) cooling at a rate of 1000-2000 K/hr.

Thus, we predict a broad range of cooling rates in the crystallization range, similar to those of Johnson et al. (2015), which are consistent with the inferred rates for CH/CB chondrules. However, and significantly, Figure 5 shows heating episodes (indicated by arrows) during cooling explained by compression of parcels of gas, caused by the expansion of neighboring parcels. These temperature fluctuations would lead to dissolution of olivine, and are required in the crystallization experiments to produce the blebby SO texture.

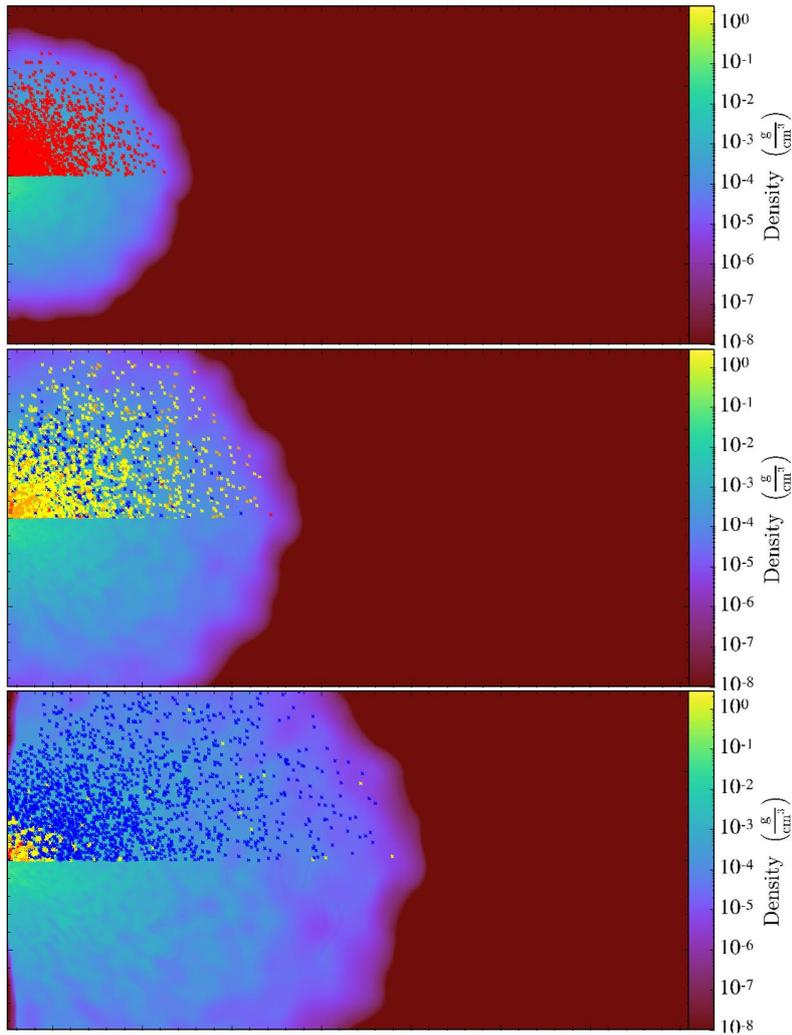

Figure 3: 3800 km wide by 1900 km high projections through the plume showing vapor density at *a)* 2 (top), *b)* 40 (middle), and *c)* 83 (bottom) minutes into the simulations. Overplotted are tracer particle locations for those particles in the upper half of the corresponding plots, which move with the flow. We only show half of the particles to allow the reader to see the structure of the plume. The tracer color corresponds to the temperature of the gas they are tracing, with red, orange, yellow, and blue corresponding to temperatures above 1800, 1600, 1400, and 0 K, respectively.

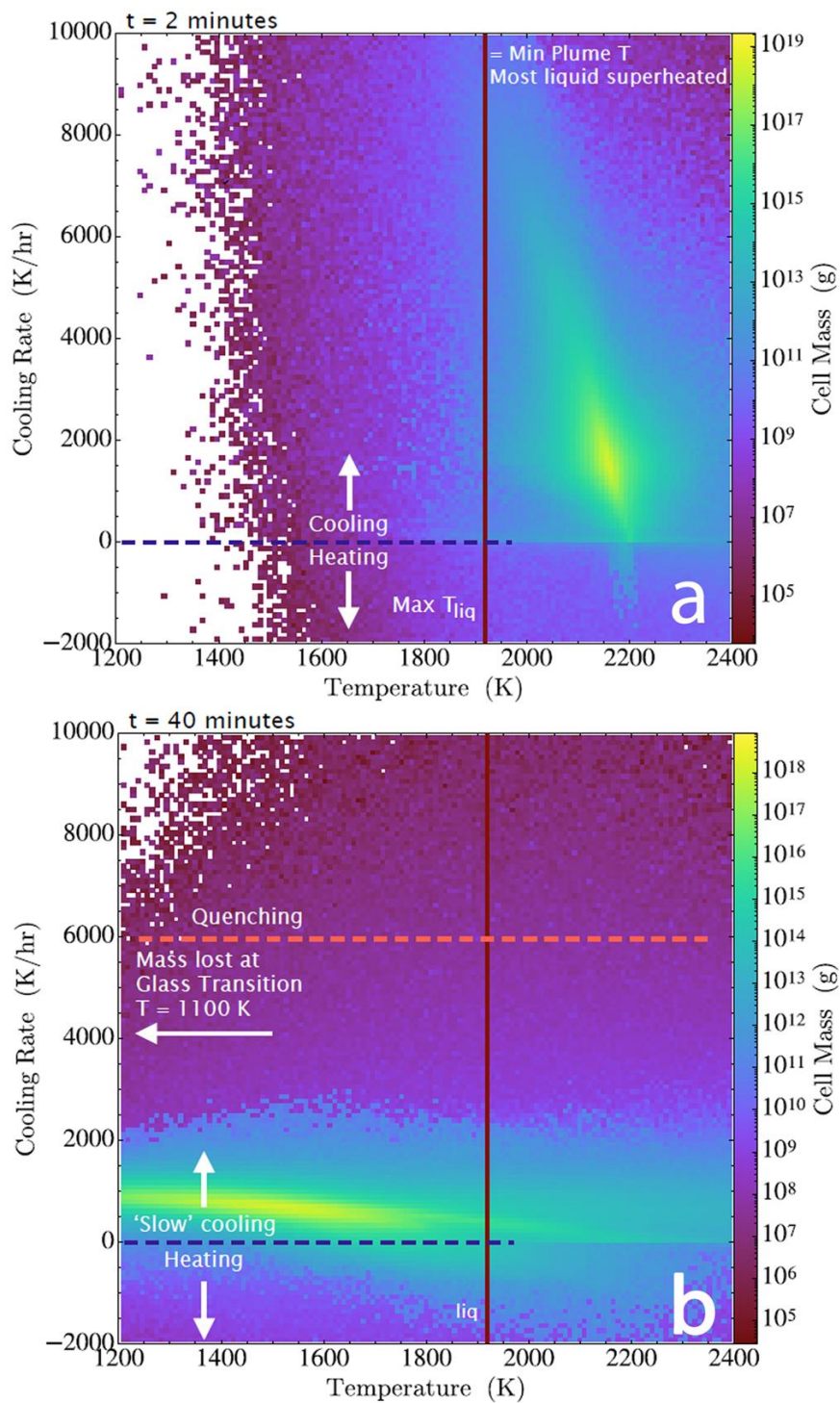

Figure 4: Ejecta fan mass distribution in cooling rate - temperature space simulated with the Eulerian AMR code FLASH4.3. (a) after 2 minutes, most of the plume (high mass region in yellow) is superheated liquid, cooling at a few thousand K/hr. (b) after ~40 minutes, most plume material is crystallizing and cooling at ~100 K/hr. In both (a) and (b) a fraction of the plume is being reheated. Fastest cooling plume material would be quenched to glass. Figures were generated using the yt analysis toolkit (Turk et al. 2011).

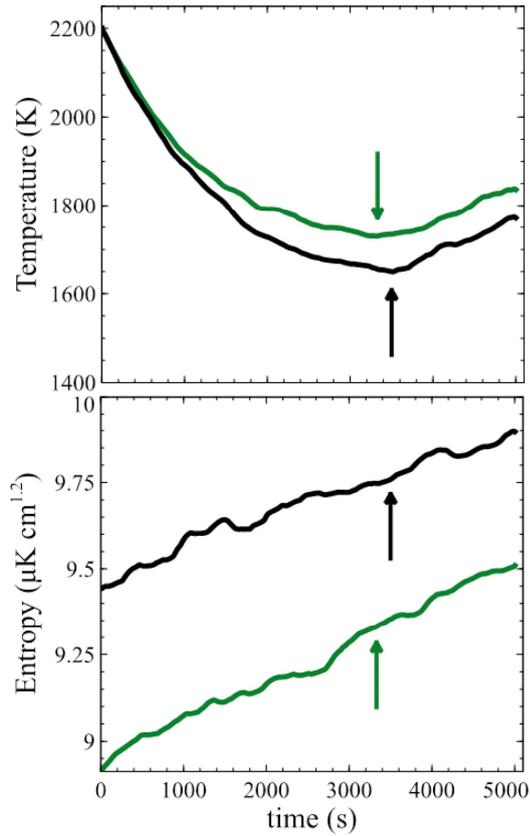

Figure 5: Temperature (top) and Entropy (bottom) profiles of different fluid parcels, smoothed with a moving average of width 27 time steps to suppress small fluctuations in tracer positions. Cooling rates are complex and vary even for a single fluid parcel. Reheating events occur (indicated by arrows) punctuate the cooling curves, which would subsequently cool over longer timescales. These heating periods correspond to periods with negative cooling rates, as shown in Figure 3. We indicate the heating transition times in the entropy plot to highlight that the transition is not due to the tracers falling into a neighboring region where the entropy value and evolution would be markedly different.

## 6. DISCUSSION

The case that metal-rich CH, $CB_a$ and $CB_b$ chondrites, part of the CR carbonaceous chondrite clan (Weisberg et al., 1995; Krot et al., 2002), contain impact plume material is strong. They have chondrules distinct in texture and composition from those in normal chondrites (Weisberg et al., 1988). Small Al-poor CC chondrules have compositions consistent with condensation after brief thermal events (Wasson, 1988; Krot et al. 2001a, 2002), as do zoned FeNi metal grains (Weisberg et al., 1988; Scott, 1988; Meibom et al., 1999; Campbell et al. 2005; Rubin et al., 2003). The Al-rich compositions of SO chondrules are complementary to those of CC chondrules, implying evaporation and condensation (Krot, 2005). Both chondrule types are enclosed by zoned condensate metal (Krot, 2001b; 2010). Pb and W isotopic data indicate that all chondrules and metal in CB chondrites formed at the same time, between 4.562 and 4.563 Ga (Krot et al., 2005; Kleine et al., 2005; Bollard et al., 2015), in the same event. The sum of these investigations leads to a consensus that a large collision produced a spray of evaporation-residue melts (SO chondrules and metal nodules) with condensation of CC chondrules and zoned metal (Krot et al., 2005, 2007).

The $CB_b$ SO chondrule textures were never found in dynamic crystallization experiments with continuous cooling paths, but appear with temperature fluctuations during cooling. Our FLASH4.3 simulation shows a wide range of cooling rates for different particles in the expanding ejecta fan (Figure 3). Despite rapid cooling at peak temperature, many "chondrules" cool at a few hundred K/hr between liquidus, solidus (~1400 K) and glass transition temperature (Figure 4). Such rates are compatible with forming normal barred olivine textures in experiments. However, temperature fluctuations appear in the model cooling paths (Figure 4) and would transform barred olivine to skeletal blebby olivine by partial remelting, as in experiments. Earlier modelling (Dullemond et al., 2014; Johnson et al., 2015) achieved reasonable cooling rates for normal chondrules, but did not model processes that could lead to heating. Specifically, they used a one-dimensional radiative cooling model, which could *only* cool, and did not self-consistently model adjacent parcels of gas in the plume and their interaction. This interaction is what leads to adiabatic cooling and *heating*, and such heating is shown by experiment to be required for the formation of $CB_b$ chondrules.

Initial modeling considered impacts of asteroid-sized bodies (Asphaug et al., 2011; Dullemond et al., 2014; Morris et al., 2015) but later work compares calculations where the target is asteroidal to those where it is a planetary embryo (Johnson et al., 2015; Hasegawa et al., 2016). The latter situation leads to higher impact velocities and therefore greater efficiency of chondrule formation. The comparison of models with different target body size shows that most of the collision-generated chondrules would have formed with the largest bodies, i.e. the final protoplanets (Hasegawa et al., 2016).

In this view, most chondrule formation could be related to planetary accretion, and one must ask if all chondrules record evidence of collisions. Clearly only the chondrules in CB and CH have unequivocal memory of a plume setting in their evidence for superheating, evaporation and condensation, dust-free accretion and especially a unique "young" age (Krot et al., 2005; Kleine et al., 2005; Bollard et al., 2015). In all other chondrules the dominant texture is porphyritic, indicating near-liquidus peak temperature; and there are surviving isotope anomalies, and relict grains. However, the abundance of the moderately volatile element Na in Type II chondrules suggests formation in a high pressure gas (Hewins et al., 2012; Fedkin and Grossman, 2013) leading to the idea that they formed in a plume, though Na is depleted in all CB chondrules. If both formed in plumes, there were collisions of very different energy for Type II and CB chondrules, lower for the older Type II chondrules. This is consistent with the results of Hasegawa et al. (2016), collisions with earlier smaller bodies producing more incompletely melted (porphyritic) chondrules. However other formation mechanisms cannot be excluded (Connolly and Jones, 2016).

7. CONCLUSIONS

CB$_b$ SO chondrules were heated to a minimum of 1928 K and experienced up to 20% evaporative loss of MgO and SiO$_2$. The smooth blebby grains of SO chondrules can be produced only with temperature fluctuations during cooling, which partially remelt barred olivine. The Adaptive Mesh Refinement impact modeling predicts a diverse range of variable cooling rates for a given fluid parcel, and thus for chondrules within that fluid parcel. Many chondrules would experience cooling at 100 K/hr for temperatures of 1400-1600K, but experience mild reheating fluctuations during their cooling history. Modeling the plume in a three dimensional mesh that resolves the plume and traces variations within it was crucial for recovering these fluctuations. Contrast this with SPH methods where the plume does not interact with the nebula, nor are pressure variations included. The thermal histories obtained by experiment and models agree, providing the first specific demonstration of the giant impact origin of the CB$_b$ chondrules. Further refinement of the model and additional parameter studies of initial conditions (e.g., different types of impacts: sizes and composition of colliding bodies, impact velocities, etc.) will be performed in future work.

We thank NASA for Cosmochemistry grants NNX14AN58G (P.I. M.A.M.) and NNX08AG62G (P.I. R.H.H.). This work used the XSEDE environment, which is supported by National Science Foundation grant number ACI-1548562. Specifically, this work used the XSEDE Stampede resource at the Texas Advanced Computing Center through allocation TG-AST160042.